\documentclass[aps]{revtex4-1}
\usepackage{graphicx}
\usepackage{bm}
\usepackage{amsmath}
\usepackage{amssymb}
\usepackage{color}
\usepackage{ulem}
\begin{document}

\title{Transition between anomalous and Anderson localization in systems with non-diagonal disorder driven by time-periodic fields}

\author{Rafael A. Molina}
\affiliation{Instituto de Estructura de la Materia - CSIC, Serrano 123, 28006 Madrid, Spain}
\author{Victor A. Gopar}
\affiliation{Departamento de F\'{\i}sica Te\'orica and Instituto de Biocomputaci\'on y F\'{\i}sica de Sistemas Complejos, Universidad de Zaragoza, Pedro Cerbuna 12, E-50009, Zaragoza, Spain}

\date{\today}

\begin{abstract}
In models of hopping disorder in the absence of external fields and at the band center, the electrons are less localized in space than the standard exponential Anderson localization. A signature of this anomalous localization is the square root dependence of the logarithmic average of the conductance on the system length, in contrast to the linear length dependence for Anderson localized systems.
We study the effect of a time-periodic external field in the scaling and distribution of the conductance of a quantum wire with hopping disorder. In the low-frequency regime, we show a transition between anomalous localization and Anderson localization as a function of the parameters of the external field. The Floquet modes mix different energy contributions and standard length dependence of the logarithmic average of the conductance is gradually recovered as we lower the frequency or increase the amplitude of the external field. In the high-frequency regime, the system presents still anomalous localization but the conductance is also renormalized, depending on the parameters of the external field, by interference effects at the coupling to the leads. This allows for a high degree of control of the average of the conductance.

\end{abstract}

\maketitle

\section{Introduction}
\label{Sec:intro}


One of the most important phenomena affecting the dynamics of waves in disordered systems is the Anderson localization: exponential decay of waves 
in space due to destructive interference of electron wave functions in a disordered medium. 
Anderson localization was originally predicted for free electrons \cite{Anderson1958} but it has been observed in both classical and quantum waves \cite{Anderson50,Chabanov2000,Roati2008,Crespi2013}. 
It is widely assumed that the presence of disorder leads to Anderson localization in 1D, however, waves can be less localized, or anomalously localized, in relation to the standard exponential Anderson localization.  

The statistical properties of the transport can reveal the presence of 
anomalous localization.  For instance, at the band center, the average of the logarithmic conductance $\langle \ln g \rangle$ depends on the system length $L$ as $\langle \ln g \rangle \propto \sqrt{L}$, in 1D disordered wires \cite{Soukoulis1981,Amanatidis2012}. This is  in contrast to the linear dependence $\langle \ln g \rangle \propto L$ expected in the presence of Anderson localization. Disordered armchair nanoribbons also show a nonlinear dependence of the average of the logarithmic conductance on $L$, near the Fermi energy \cite{Kleftogiannis2013}. Similarly, disorder models characterized 
by the so-called L\'evy distributions, lead softer localization properties than Anderson's as has been theoretically 
\cite{Falceto2010,Amanatidis2012} and experimentally \cite{Barthelemy2008,Fernandez2014} explored. In all these examples \cite{Soukoulis1981,Falceto2010,Amanatidis2012,Kleftogiannis2013,Barthelemy2008,Fernandez2014}, quantum (electrons) and classical (electromagnetic) waves are anomalously localized.

Quantum control using external time-periodic fields, sometimes referred to as Floquet engineering, is a rapidly growing field due to its potential use for applications and for controlling and expanding the properties of new materials. One of the first theoretical examples of this kind of control came from one-dimensional disordered systems \cite{Holthaus1995,Holthaus1996}. The localization length in a disordered quantum wire can be controlled by coherent light linearly polarized in the growing direction of the wire \cite{Martinez2006a,Martinez2006b}. Later, the conductance distribution of an ensemble of disordered wires with driving ac fields was studied \cite{Gopar2010}. It was shown that the results in the high-frequency regime could be explained by the distribution without driving with a change in the localization length and a renormalization factor. In the low frequency regime, the changes in the conductance distribution could be interpreted as due to the increase in the number of channels available for transport as the systems exchange photons with the external laser field \cite{Gopar2010,Kitagawa2012}. 
The high-frequency regime usually allows for analytical approximations or perturbation expansions whose result is a typical renormalization of parameters with very interesting possibilities of control like, for example, a transition between trivial and topological behavior in Floquet topological insulators or Floquet topological semimetals\cite{Lindner2011,Cayssol2013,Wang2014,Hubener2017,Gonzalez2017}. The 
low-frequency regime can usually be treated only through numerics but presents a richer behavior and new physics \cite{GomezLeon2013,Martiskainen2015,Gonzalez2016,Benito2017}. Although most realistic low-dimensional experimental systems need to cope with the presence of disorder, results for Floquet control in the presence of disorder are still few. Apart from the previously mentioned one-dimensional wires, there has been some interesting results in Floquet-Anderson insulators \cite{Titum2016}, in controlling the localization of interacting bosons and the many-body localized phase \cite{Santos2009,Ponte2015,Abanin2016}, in the response to multifrequency driving \cite{Hatami2016}, and studying the different response regimes of periodically driven Anderson insulators \cite{Liu2018}.


In this paper, we study the effect of external fields in disordered one-dimensional systems with anomalous localization. We focus on a simple model for anomalous localization defined in a lattice with disorder in the hopping between neighboring sites. The high-frequency regime usually appears for frequencies larger than the energy span of the density of states. The long tails in the density of states that appear in the case of non-diagonal disorder imply that the high-frequency regime is strictly valid only for very high values of the frequency. However, the transition between the low-frequency regime and the 
high-frequency regime is much smoother than in the case of diagonal disorder and, in fact, this was one of the motivations for our study. As a matter of fact, we have seen that the density of states decays rapidly for values of the energy far from the band center and there is a certain range of parameters for which the 
high-frequency approximation works very well even if the frequency is not as high. In this regime, the possibilities for controlling the conductance of the system are very clear as the values of each individual system are modified by a common factor related to the phenomena of coherent destruction of tunneling \cite{Gopar2010}. The low-frequency regime is more difficult to interpret. The conductance distributions do not change from the case without external field up to a certain value of the length of the system dependent on the frequency and amplitude of the driving field. After that, the exponent $\alpha$ characterizing  the anomalous localization changes and for large lengths, the standard exponential localization scaling is recovered. The conductance distributions show a transition between the anomalous regime of localization and the standard exponential regime as a function of reducing the frequency or increasing the amplitude of the external field. We attribute this transition to the mixing of the contributions of different energies due to the increase in the weight of higher order Floquet modes. 

The outline of the rest of the paper is as follows: First, we describe the model in Sec. \ref{Sec:model}, then we present the results in the high-frequency regime in Sec. \ref{sec:highfreq} and the results in the low-frequency regime in Sec. \ref{sec:low}. The conductance distributions in the transition between the anomalous localization regime and the standard localization are discussed in Sec. \ref{sec:tran}. The paper is finished with conclusions in Sec. \ref{Sec:conclusions}

\section{Random hopping model with driving}
\label{Sec:model}
We use the one-dimensional Anderson model with nearest neighbor random hopping described by the Hamiltonian
\begin{equation}
\label{eq:Hnot}
H_0= \sum_n t_n ( c_{n}^\dagger c_{n+1} +c_{n+1}^\dagger c_{n} ) ,
\end{equation}
where $c_n^\dagger$ and $c_n$ are creation and annihilation operators for spinless fermions, and $t_n$ are the random hopping  elements sampled from a distribution of the  form $P(t)=1/wt,
\exp(-w/2) \leq t \leq \exp (w/2)$, where $w$ denotes the strength of the disorder. This is the so-called logarithmic off-diagonal disorder. \cite{Soukoulis1981} As we have mentioned, the model described by Eq. (\ref{eq:Hnot}) has been found to present unconventional localized states at zero energy  \cite{Soukoulis1981,Ziman1982} whereas, for nonzero energy, standard localized states are present. This phenomenon has been studied extensively in the literature \cite{Soukoulis1981,Ziman1982,Inui1994,Brouwer1998,Evangelou2003}.

\begin{figure}
\centering
\includegraphics[width=0.9\columnwidth]{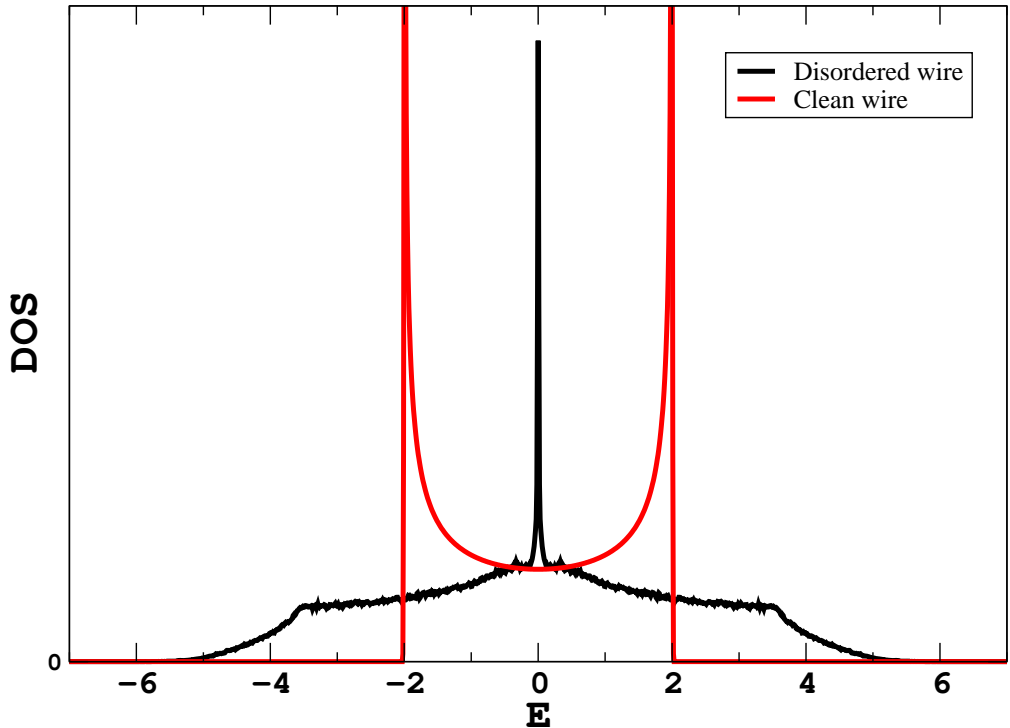}
\caption{Density Of States(DOS) in arbitrary units as a function of the energy for the Anderson model with exponential hopping disorder described by Eq. \ref{eq:Hnot} and a model for a clean wire.}
\label{fig:dos}
\end{figure}

\begin{figure}
\centering
  \includegraphics[width=0.9\columnwidth]{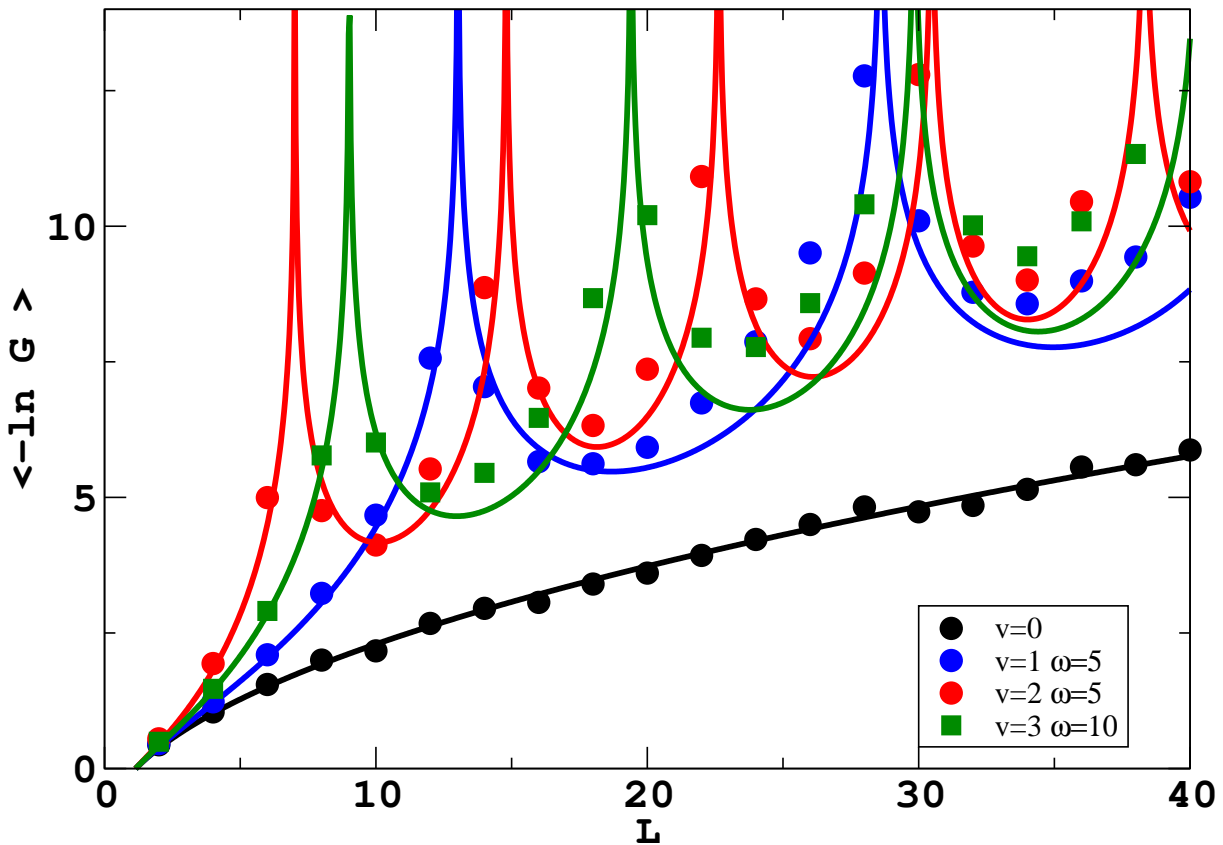}
\caption{Results for the average logarithmic conductance as a function of the length for the case without driving at $E=0$ and for three different examples in the high frequency regime. The solid lines are the analytical results from Eq. \ref{eq:avlg_highf}, while the black solid line shows the power-law behaviour: $\langle \ln G \rangle \propto L^{1/2}$.}
\label{fig:highf}       
\end{figure}

In Fig. \ref{fig:dos} we plot the Density Of States (DOS) as a function of the energy for an ensemble average of 1000 disordered wires with the disorder strength $w=2.5$ and number of sites $L=1000$ and compare with the DOS for a clean wire of the same length with all hopping matrix elements equal to one. In the disordered case, the abrupt transition of the DOS at $|E|=2$ disappears and we can see how it acquires long tails which extend noticeably up to $E \approx 6$. The actual DOS is only zero for energies larger than twice the maximum hopping allowed by the disorder distribution which for the disorder strength chosen is $|E|=6.9807$. The shape of the DOS will be very important later to understand the transition between the high and low-frequency regimes in the disordered case with external time-periodic field. The peak at zero energy marks the position of the anomalously localized states. The length dependence of the average logarithmic conductance $\left< \ln g \right> \propto L^{1/2}$ without driving at zero energy, 
is shown as the black curve in Fig. \ref{fig:highf} and compared with numerical results (black dots). 

To the static Hamiltonian of Eq. \ref{eq:Hnot} we add a periodic driving:
\begin{equation}
H(t)=H_0+\sum_n v na \cos{(\omega t)} c_{n}^\dagger c_{n}.
\label{eq:driving}
\end{equation}
In the electronic case, this driving corresponds to laser light of frequency $\omega$ and amplitude $v$ linearly polarized in the growing direction of the wire. The parameter $a$ is just the distance between sites in the tight-binding description that we take $a=1$. Please, note that this linear potential corresponds to a constant oscillating electric field. This formulation is equivalent via a gauge transformation to a time-dependent modulation of the hoppings. In this paper, we use natural units with $\hbar=1$ but, in the following, we will recover the explicit dependence on $\hbar$ when it helps clarify the physical meaning of the formulas.


We are interested in the transport properties of the driven wires as they are directly related to the localization properties of the electronic states. We connect the one-dimensional wires to left and right leads and 
define the dc-conductance for an ac-driven quantum wire as
\begin{equation}
G=\lim_{V \rightarrow 0} \frac{d \bar{I}}{dV} ,
\end{equation}
where $\bar{I}$ is the current averaged over one period of the driving field.
Thus $G$ is an experimentally accessible quantity which can be written as a sum of the Floquet modes
$g^{(k)}(E_F)$:
\begin{equation}
\label{G}
 G=\sum_{k=-\infty}^\infty g^{(k)}(E_F) ,
\end{equation}
where
\begin{equation}
\label{gk}
g^{(k)}(E_F)= \frac{1}{2} \left[ T_{1N}^{(k)}(E_F) + T_{N1}^{(k)}(E_F) \right] ,
\end{equation}
with $T_{1N}^{(k)}$ being the transmission for electrons from the left to the right lead, similarly   $T_{N1}^{(k)}(E)$ for electrons from the right to the left. In the static case, $T^{(k)}_{1N} = T^{(k)}_{N1}$
 for systems with time-reversal symmetry, however, in the presence of time-dependent fields, this is no longer true \cite{Kohler2005}. When both transmissions are not equal, a current at zero voltage can be induced by the external field. This pumped current appears as an offset at zero voltage in the $I\mathrm{-}V$ characteristic of our system. The presence of a pumped current does not modify our formulas, as we calculate the slope of the $I\mathrm{-}V$ curve at zero
voltage. In the disordered systems, we considered the pumped current is zero on average and as the length of the systems
increases is, in general, negligible. We will, then,  ignore it in our calculations.

The transmissions needed for the generalization of the Landauer formalism to the Floquet case can be obtained from the Floquet-Green functions. As the Hamiltonian is time-periodic with period $T=1/\hbar\omega$, the Floquet theorem states that
there is a set of solutions $|\phi_\epsilon^{n, m}(t)\rangle$ to the equation
\begin{equation}
\label{floquetequation}
 \left[H(t)-i\hbar \frac{d}{dt}\right]|\phi_\epsilon^{n, m}(t)\rangle =\epsilon^{n, m}|\phi_\epsilon^{n, m}(t)\rangle ,
\end{equation}
where  $\epsilon^{n, m}=\epsilon_n+m\hbar\omega$ with $m$ an integer number and $n$ the index of the states in the first Floquet-Brillouin zone: $-\hbar\omega/2 \le (\epsilon_n) \le \hbar \omega/2$.
The Green's function $ G(E,t',t'')$ satisfying $\left[ {\mathbb I} E - H(t') \right]G(E,t',t''){\mathbb I}={\mathbb I}\delta_T(t'-t'')$, where ${\mathbb I}$ is the identity operator and $\delta_T(t)$ is a $T$-periodic delta function, can be written in terms of the Fourier components of the Floquet
states $|\phi_m^{n,0}\rangle = \frac{1}{T}\int^T_0 e^{im\omega t'} | \phi^{n,0}(t')\rangle dt'$. Thus the Fourier components $G^{(k)}(E)$ of $G(E,t',t'')$ can be written as
\begin{equation}
\label{Gk}
G^{(k)}(E)=\sum_{n,m}\frac{|\phi_{k+m}^{n, 0}\rangle \langle {\phi_m^{n, 0}}|}{{E-\epsilon_n-m\hbar\omega}} .
\end{equation}
From the Floquet Green functions we can obtain the transmissions needed to calculate the conductance \cite{Martinez2003,Martinez2005,Kohler2005}:
\begin{equation}
T_{1N}^{(k)}(E)=\Gamma_1\Gamma_N|G_{1N}^{(k)}(E)|^2,
\end{equation}
and
\begin{equation}
T_{N1}^{(k)}(E)=\Gamma_N\Gamma_1|G_{N1}^{(k)}(E)|^2.
\end{equation}
The parameters $\Gamma_N$ and $\Gamma_1$ are respectively the coupling constants to the right lead, which is coupled to the last site of the wire, and the left, coupled to the first site of the wire. We assume perfect coupling to the leads and take these constants as unity, including the effect of the leads as a constant self-energy in the Green's function. This wide band limit approximation implies that the coupling to the leads is constant at all energies relevant for the transport properties. This is usually a good approximation for metallic leads.

We are thus interested in the statistics of the conductance $G$ given by Eq. (\ref{G}). The conductance distributions are found numerically by sampling over different disorder realizations, i.e., over different configurations of $t_n$ in Eq. (\ref{eq:Hnot}) and we assume perfect lead-wire coupling. The numerical simulations are performed according to Eqs. (\ref{Gk}) and (\ref{G}), where the Fourier components $G^{(k)}$ are calculated using the method of matrix continued fractions. \cite{Medina2001,Martinez2003,Martinez2005,Martinez2006a,Martinez2008}. This is a very demanding task numerically as the number of Floquet modes needed for convergence is quite large, increasing for larger wires. For example,  the 
low-frequency results shown in Fig. \ref{fig:lowf} around 240 modes were needed for convergence of the larger wires used. The histograms shown in the figures of this work were obtained from 10000 different disorder realizations. However, due to time limitations, the average values of the conductance as a function of the length of the system were obtained with 1000 realizations.

\section{High frequency: coherent destruction of tunneling}
\label{sec:highfreq}
The high-frequency regime is usually characterized by a frequency larger than the bandwidth of the system. The bandwidth for the disordered hopping model is twice the maximum value of the hopping allowed by the disorder. They have taken the value of the disorder strengths in the simulations as $w=2.5$, so the simplest expectation is that the high-frequency regime will set in for frequencies $\hbar \omega > 4e^{(2.5/2)} \approx 13.96$. However, as we have seen in the previous section, the density of states is quite low at energies $|E|<4$ so the analytical results for high frequency are approximately valid for lower frequencies than one would naively expect.

It is well established since the early works of Dunlap and Kenkre and Grossmann {\it et al.} that, in the high frequency regime, the driving system behaves as an autonomous system with a renormalized hopping term \cite{Dunlap1986,Grossmann1991}
\begin{equation}
t_{eff}=t J_0 \left(\frac{va}{\hbar\omega}\right).
\label{eq:cdt}
\end{equation}
This coherent destruction of tunneling occurs in clean disordered systems alike and is responsible for the renormalization of the localization length in the Anderson model with diagonal disorder and time-periodic driving \cite{Holthaus1995,Holthaus1996,Martinez2006a}.
However, in the non-diagonal disorder case a global factor renormalizing the hopping terms does not modify the localization length of the system as the relative disorder strength is unchanged.
In spite of the unchanged value of the localization length, the conductance in a two-terminal setup is modified by the coherent destruction of tunneling in the coupling between system and leads. In the high-frequency regime in the absence of disorder, the conductance is proportional to $J_0^2(vL/\hbar\omega)$ \cite{Martinez2008}, with $L$ equal to the total length of the system. This factor remains under the presence of disorder, independently on whether it is diagonal or non-diagonal \cite{Gopar2010}. The implications for the length dependence of the average logarithmic conductance are clear:
\begin{equation}
\label{eq:avlg_highf}
\left<\ln G \right>=\left< \ln g \right > - 2\ln|J_0(vL/\hbar\omega)|,
\end{equation}
where $g$ is the conductance of the system without driving.

In Fig. \ref{fig:highf} we show a comparison between the results of the average logarithmic conductance between the case with no driving and disorder $w=2.5$ and three different examples of driving in the high-frequency regime. The black solid lines are a fit to the case without driving of a power-law length dependence with an exponent corresponding to anomalous localization of $1/2$. The colored solid lines correspond to the results of Eq. \ref{eq:avlg_highf}. The agreement with the theoretical result is quite good even for the two cases with frequencies $\omega=5$ which is outside the strict high-frequency limit. The agreement fails for larger systems as the high-frequency renormalization of the coupling terms is a first order perturbation result in the amplitude of the field over the frequency and we expect higher orders in perturbation theory to be relevant as the length of the system increases. In Sec. \ref{sec:tran} we will see that not only the average is modified by the field-dependent factor but the entire distribution is the distribution for anomalous localization displaced. In Figs. 2 and 3, we have only used data for even number of sites as even-odd oscillations of the conductance occurring for clean wires may survive the presence of disorder for shorter wires \cite{Yamaguchi1997,Molina2004,Kwapinski2005,Martinez2008}.

\section{Low frequency: transition between anomalous and standard localization}
\label{sec:low}

\begin{figure}
\centering
  \includegraphics[width=0.9\columnwidth]{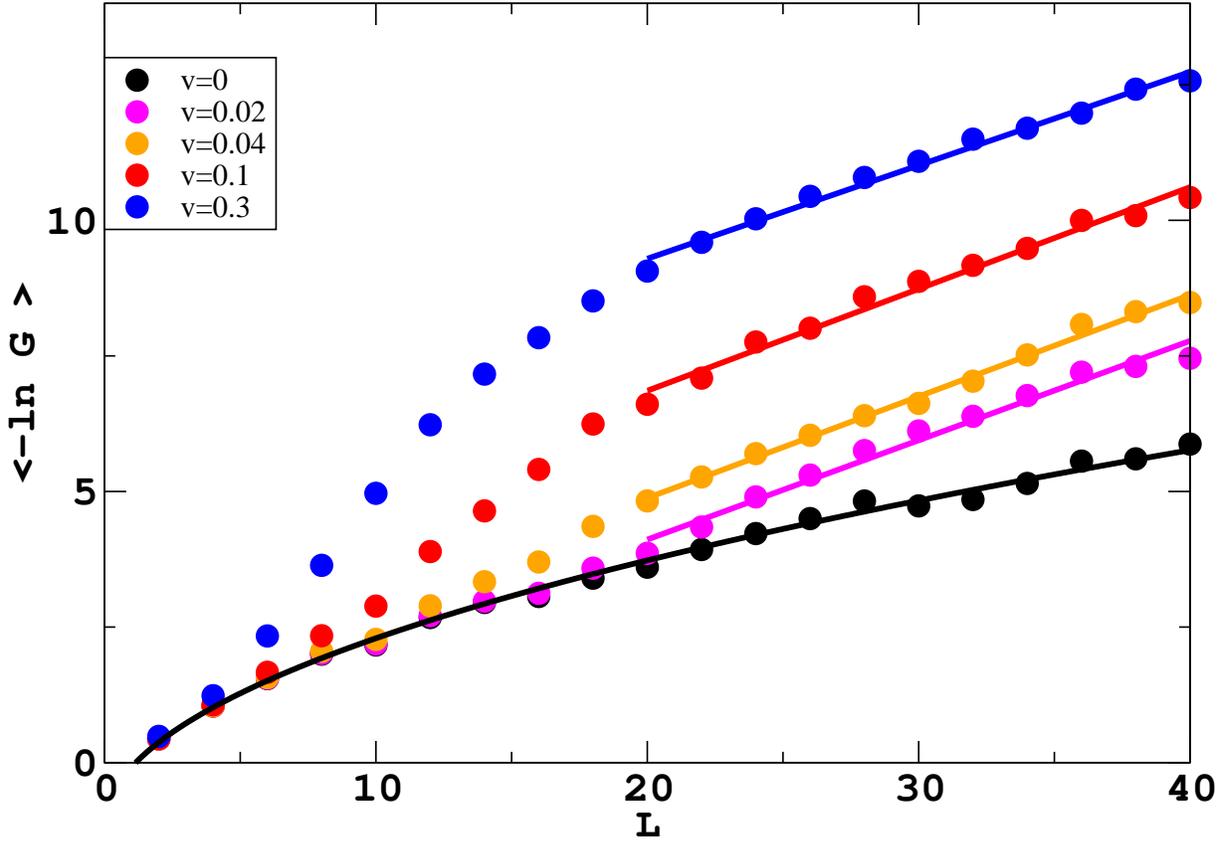}
\caption{Average logarithmic conductance $\left< ln G \right>$ vs. system length $L$ for several values of the amplitude of the field and frequency $\omega=0.2$ well into the low frequency regime. The straight lines for $L>20$ are linear fits as expected to the results for standard exponential localization for values of $L>20$, while the black line corresponds to the power-law behaviour $\langle \ln G \rangle \propto L^{1/2}$.}
\label{fig:lowf}      
\end{figure}

In Fig. \ref{fig:lowf} we show the results for the average logarithmic conductance as a function of the length of the system for several values of the amplitude of the field $v$ with frequency $\omega=0.2$, including the no driving results with $v=0$. The results for shorter systems are equal to the results without driving. However, as the length of the system is increased, the length dependence changes and moves away from the expected power-law dependence of anomalous localization to the linear behavior for $L>20$ expected in the case of standard Anderson localization. 
This transition occurs due to the mixing of different energy contributions in the Floquet modes as the particles exchange photons with the external field. A certain minimal length depending on the amplitude is needed for the Floquet modes outside the zero mode to start being non-negligible. As the zero energy is the only energy with anomalous localization in our model, when the Floquet modes outside the zero one are important we get a transition to the standard localization that characterizes energies outside the middle of the band. 

\section{Conductance distributions}
\label{sec:tran}

To study the full conductance distributions in the presence of the driving field for our model with anomalous localization, we first briefly present a main result in Refs. \cite{Falceto2010,Amanatidis2012,Kleftogiannis2013} 
to describe  the statistical properties of the conductance for anomalously 
localized electron wavefunctions for no external driving fields applied. After that, we extend the analysis to external fields. 
It was found that the distribution of conductances $P_{\alpha, \xi}(g)$ is determined by two parameters: the power $\alpha$ of the dependence of the average $\langle \ln g  \rangle$ with the length $L$, i.e. $\langle \ln g \rangle \propto L^\alpha$ and the value $\xi =\langle \ln g \rangle$. With the knowledge of these two quantities the 
distribution $P_{\alpha, \xi}(g)$ is given by
\begin{equation}
\label{pofg_xi}
P_{\alpha,\xi}(g)=\int_0^\infty p_{s(\alpha,\xi,z)}(g) q_{\alpha,1}(z){\rm d}z ,
\end{equation}
where 
\begin{equation}
s(\alpha,\xi,z)={\xi}/(2{z^\alpha I_\alpha)}, 
\end{equation}
with $I_\alpha =1/2 \int_{0}^{\infty} z^{-\alpha} q_{\alpha,1}(z)dz$. $p_{s(\alpha,\xi,z)}(g)$ in Eq. (\ref{pofg_xi}) is the probability density distribution for the case of standard Anderson localization, which is known within a single parameter scaling approach \cite{Mello2004}, while $q_{\alpha,1}$ is the 
probability density function of the L\'evy-type distribution supported in the positive semiaxis  with parameter $\alpha$ and scale parameter equal to unity.

We now consider that an external periodic field is applied. In the high frequency limit  we have seen that coherent destruction of tunneling implies:
\begin{equation}
\label{Gproptog}
G = J_0^2\left(\frac{v L}{\hbar\omega}\right) g ,
\end{equation}
where we have used $G$ for the adimensional conductance in the presence of driving and $g$ for the same quantity in the autonomous system. The probability $P_s(G)$ is thus given by \cite{Gopar2010}
\begin{eqnarray}
\label{pofGapprox}
&& p_s(G)=\frac{C_G}{G^{3/2}}\frac{J_0^{3/2}(\nu)}{(J_0^2(\nu)-G)^{1/4}} \sqrt{\mathrm{acosh}{\left({J_0(\nu)}/{\sqrt{G}}\right)}} \nonumber \\ &\times&
\exp{\left[ -s^{-1}\mathrm{acosh}^2\left({J_0(\nu)}/{\sqrt{G}}\right) \right]} ,
\end{eqnarray}
where $\nu=vL/\hbar\omega$, $C_G$ is a normalization constant and $s \equiv L/l$, which can be  obtained from Eq. (\ref{Gproptog}) as
\begin{eqnarray}
\label{newlocalization}
{L}/{l} &=&2  \ln J_0(\nu)-\langle \ln G \rangle, 
\end{eqnarray}
where we have used that $L/l=\langle -\ln g \rangle$.

We now combine the above results (Eqs. \ref{pofg_xi} and \ref{pofGapprox}).
As we have pointed out  (see the black curve in Fig. \ref{fig:highf}), at the band center and no driving field $\langle \ln g \rangle \propto L^{1/2}$. We thus identify  $\alpha=1/2$ and write
the distribution of the conductance at the band center with an applied driving field, in the high-frequency regime as
\begin{eqnarray}
\label{pofG_xi}
P_{1/2,\xi}(G)=\int_0^\infty p_{s(1/2,\xi,z)}(G) q_{1/2,1}(z){\rm d}z ,
\end{eqnarray}
where $p_{s(1/2,\xi,z)}(G)$ is given by Eq. \ref{pofGapprox} with $s$ replaced by 
$s(1/2,\xi,z)=\xi/(2z^{1/2}I_{1/2})$. In general, there is no  analytical expression for $q(\alpha,c)$, but for the particular case of $\alpha=1/2$ (the so-called L\'evy distribution), which is the one of interest for our model with hopping disorder, it is given by
\begin{equation}
 q_{1/2,1}(z)=\frac{1}{\sqrt{2 \pi}}z^{-3/2}e^{-1/2z}.
\end{equation}

\begin{figure}
\centering
  \includegraphics[width=0.9\columnwidth]{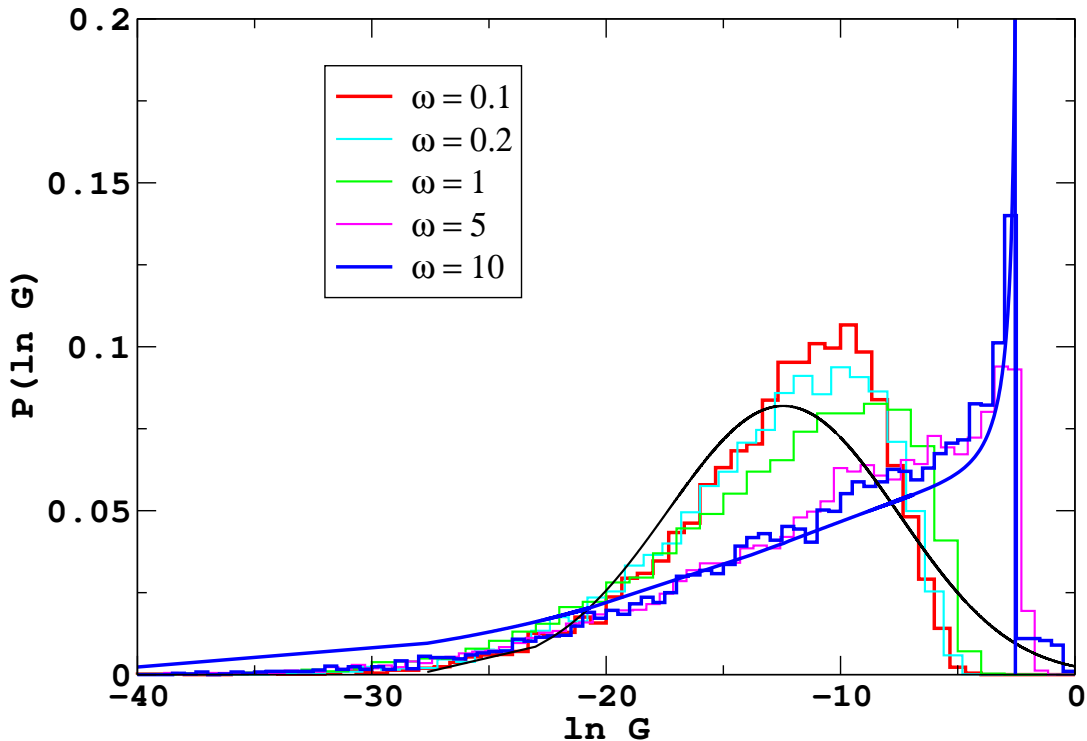}
\caption{Full conductance distributions in the transition between anomalous and standard localization.}
\label{fig:distribution}       
\end{figure}

In Fig. \ref{fig:distribution}, we show how the conductance distribution evolves for the same field amplitude $v=2$ as we lower the frequency. The blue solid line has been obtained by numerical integration according to Eq. (\ref{pofG_xi}). In the high-frequency cases $\hbar\omega=10$ and $\hbar\omega=5$, the distribution is well described by the previous theory, although the cutoff factor due to coherence destruction of tunneling is not strictly due to the long tails of DOS described in Sec. \ref{Sec:model}. 

As the frequency gets lower, the shape of the distribution changes dramatically and it appears closer to a log-normal distribution  that characterizes the fully localized regime of standard localization in one-dimension. A log-normal distribution with mean value $\langle -\ln G \rangle=12.8$, which corresponds to the lowest value $w=0.1$ is shown as the black curve in Fig. \ref{fig:distribution}. 
However, we do not fully attain the deep localized limit. We believe that the correct distribution should be fitted by the multichannel quasi-1D case as was shown before in the low-frequency regime of the diagonal Anderson model\cite{Gopar2010}. However, the contributed number of channels is very high (around 40 for $\hbar\omega=1$ so the calculation of the multidimensional integrals involved is outside our numerical capabilities \cite{Gopar2002,Muttalib2003}.      

\section{Conclusions and perspectives}
\label{Sec:conclusions}

In conclusion, we have studied the scaling of the average conductance and the conductance distribution of disordered one-dimensional wires with non-diagonal disorder subject to time-periodic driving with a constant force along the direction of the wire. In the absence of driving such a system presents anomalous localization at the band center. 

When driving is included, there are two distinct regimes. In the high-frequency regime, the conductance distribution is renormalized by a factor that depends on the coherence destruction of tunneling of the coupling to the leads. Although the distributions can be obtained from the distributions of the undriven systems, the behavior of the average conductance with length depends on complex interference effects and is very non-monotonous. In that case, even in the presence of disorder, the average conductance can be controlled through the parameters of the external field. 

On the other hand, in the low-frequency case we observe a transition from anomalous localization to standard localization as the frequency is reduced or as the amplitude of the field is increased. The conductance distributions separate from the anomalously localized case and become similar to a log-normal distribution, which is characteristic of Anderson localized wires. However, we do not have a complete theory 
that can describe the transition from high to low-frequency limits of the conductance statistics. 
In the low-frequency regime, each Floquet mode acts as a different channel although the 
conductance is limited by the actual number of channels and these channels are not fully uncorrelated. We conjecture. that these differences are related to a multichannel effective description which due to the high number of channels involved is outside our current capabilities to calculate.

Our conclusions for the high-frequency case are very compelling. Although the system sizes that is possible to reach to study the conductance distribution with Floquet theory are quite limited, we are convinced of the merit of our interpretation of the low-frequency results.
Experimentally, the system studied here may model a quantum wire illuminated by laser light. We also expect our results to encourage more theoretical and experimental studies on the interplay between periodic perturbations and disorder. In this sense, it would be very interesting to look for similar effects in different systems where anomalous localization has been found like disordered graphene nanoribbons or systems with Levy-type disorder.

\section{Acknowledgments}
Financial support through Spanish grants
PGC2018-094684-B-C22 and 
PGC2018-094180-B-I00 (MCIU/AEI/FEDER, EU), FIS2015-63770-P (MINECO/FEDER, EU), CAM/FEDER Project No.S2018/TCS-4342 (QUITEMAD-CM) and CSIC Research Platform PTI-001 is gratefully acknowledged.

\section{Authors contributions}
All the authors contributed equally to this work.
All the authors were involved in the preparation of the manuscript.
All the authors have read and approved the final manuscript.
%
%

\end{document}